# THE SPIN-BARRIER RATIO FOR S AND C-TYPE MAIN ASTEROIDS BELT

Albino Carbognani

Astronomical Observatory of the Aosta Valley Autonomous Region (OAVdA)

Lignan 39, 11020 Nus (Aosta), ITALY

albino.carbognani@gmail.com

## Abstract

Asteroids of size larger than 0.15 km generally do not have periods $P$ smaller than about 2.2 hours, a limit known as cohesionless spin-barrier. This barrier can be explained by means of the cohesionless rubble-pile structure model. In this paper we will explore the possibility for the observed spin-barrier value to be different for C and S-type Main Asteroids Belt (MBAs). On the basis of the actual bulk density values, the expected ratio between the maximum rotation periods is $P_C/P_S \approx 1.4 \pm 0.3$. Using the data available in the asteroid LightCurve Data Base (LCDB) we have found that, as regards the mean spin-barrier values and for asteroids in the 4-20 km range, there is a little difference between the two asteroids population with a ratio $P_C/P_S \approx 1.20 \pm 0.04$. Uncertainties are still high, mainly because of the small number of MBAs with known taxonomic class in the considered range. In the 4-10 km range, instead, the ratio between the spin-barriers seems closer to 1 because $P_C/P_S \approx 1.11 \pm 0.05$. This behavior could be a direct consequence of a different cohesion strength for C and S-type asteroids of which the ratio can be estimated.

**Keywords**: asteroids, spin-barrier, asteroids rotation

## Introduction

Asteroids in the Main Belt (located roughly between the orbits of Mars and Jupiter), were subject to strong collisional interaction and the population that we see today is the result of billions of years of evolution. Pravec and Harris (2000), in their classical analysis on rotation periods, have argued that objects with size larger than about 0.15 km do not have periods smaller than about 2.2 hours (cohesionless spin-barrier), see Fig. 1. This spin barrier can be explained by means of a cohesionless "rubble-pile" structure model, in which asteroids with diameter greater than or equal to about 0.15 km are made up of collisional breakup fragments bound together by mutual gravitational force only. The asteroids with smaller diameters are considered instead as monolithic blocks, i.e. collisional fragments rotating faster than the spin-barrier value because of the strong internal solid-state forces that hold the body together. Rotation with periods exceeding this critical value will cause asteroid breakup and the formation of a binary system (Pravec and Harris, 2007).

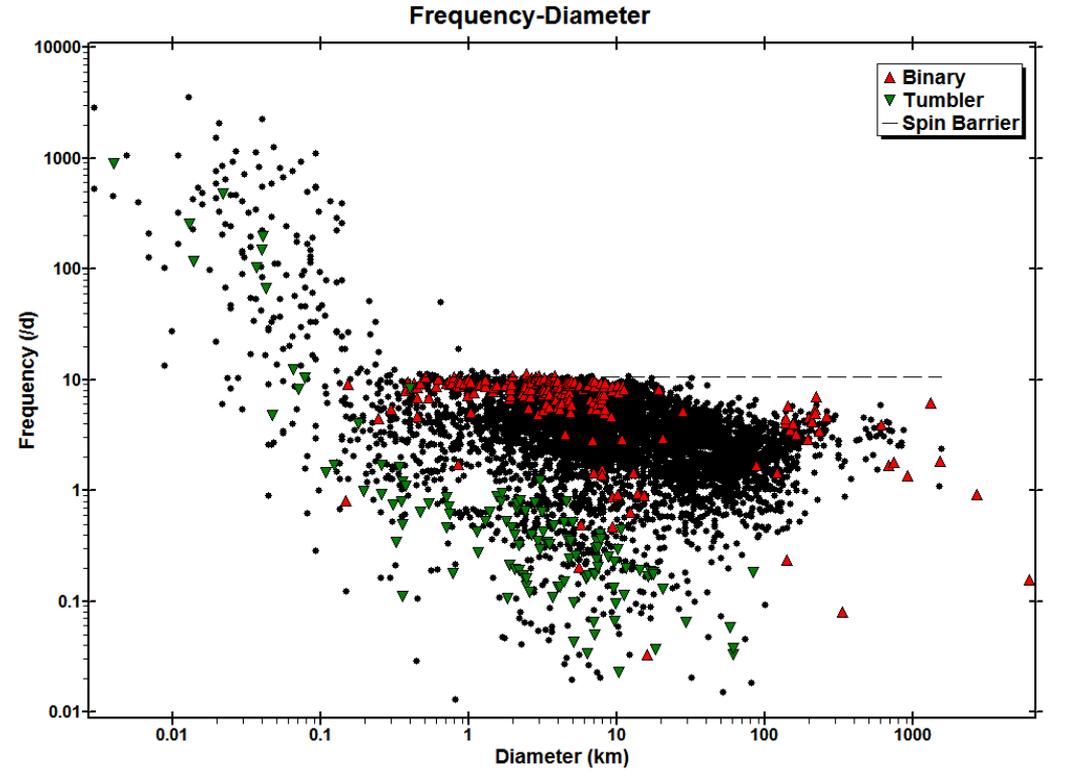

Figure 1. The periods of minor planets plotted as frequency (rotations/day) vs. diameter (km). This figure was drawn from the Asteroid Lightcurve Photometry Database (alcdef.org), held by Brian D. Warner. The spin-barrier threshold is evident for asteroids with diameter between about 0.15 - 20 km.

In a cohesionless rubble-pile structure model it is the balance between gravitational and centrifugal forces that determines the value of the critical rotation period. For a spherical asteroid, this critical period is given by (Pravec and Harris, 2000):

$$P_{crit} = \sqrt{\frac{3\pi}{G\rho}} \approx \frac{3.3 \text{ h}}{\sqrt{\rho}} \qquad (1)$$

In Eq. (1) $G$ is the gravitational constant and $\rho$ is the bulk density expressed in g/cm$^3$. For example, with $\rho$ = 2.5 g/cm$^3$, we found $P_{crit} \approx$ 2.1 h. As we can see, the critical period depends on the bulk density: the greater is the density the smaller is the rotation period of the rubble-pile. So if we have two populations of asteroids with different bulk density it is reasonable to expect a difference in the spin-barrier values. In our Solar System this situation occurs for C and S-type asteroids in the Main Belt. The majority of MBAs, have semi major axes within the range 2.2–3.2 AU (Astronomical Unit). The S-type asteroids are located near the inner edge of the Main Belt, while the C-type are near the outer edge (Carbognani, 2010). Generally speaking, C asteroids are more difficult to be observed than S ones, because they are at a greater mean distance from the Sun and show a lower mean geometric albedo. Unfortunately, the bulk density of these two populations is not well-known.

## The density of asteroids and the spin-barrier ratio

The asteroid's bulk density is a crucial but difficult parameter to obtain, as we need to know both the body mass and the volume. Data on the bulk density of asteroids have increased in the last years and have led to significant insights into the structure of these objects. The advancement of knowledge comes mainly from observations of asteroid satellites, asteroid mutual gravitational events, perturbations on neighbouring spacecraft and dynamical models of perturbing acceleration on Mars. However, there are less than 300 density estimates, so the measures are still very limited compared to the asteroids number. For a well written and recent review about asteroids bulk density estimates see Carry, 2012. As far as we are concerned, we are only interested in the bulk density of C and S-type asteroids. In general S-type asteroids are more dense than C-type, which probably have larger macroporosity. Furthermore, the bulk density, both for the S and C-types (but especially for the latter), seems to increase with the mass, probably owing to decreasing porosity (Carry, 2012).

The best bulk density estimate is $\rho_S = 2.72 \pm 0.54$ g/cm$^3$ for S-type and $\rho_C = 1.33 \pm 0.58$ g/cm$^3$ for C-type. In the first case the relative error is $\pm$ 20%, in the second is $\pm$ 44%. So the mean ratio is about $\rho_S/\rho_C \approx 2.0 \pm 1$ (Carry, 2012; DeMeo and Carry, 2013). Much of the uncertainty of the previous ratio came from the fact that on the bulk density of C-type as the relative error is greater. With these best values, from Eq. (1), the expected ratio between the spin-barrier is:

$$\frac{P_C}{P_S} = \frac{\omega_{0S}}{\omega_{0C}} = \sqrt{\frac{\rho_S}{\rho_C}} \approx 1.4 \pm 0.3 \qquad (2)$$

In Eq. (2) $\omega = 24/P$, with $P$ in hours, is the rotation frequency expressed in rot/day. The relative error is about 20%. To reduce it, a more accurate estimate of the density ratio between C and S is required. In our case we assume that the bulk density is mass independent but, especially for small diameters, it seems that the bulk density of C-type asteroids may be below 1 g/cm$^3$ (Carry, 2012). In this case the previous spin-barrier ratio may also be higher.

In this regard it is interesting to note that a rotating frequency analysis we made previously showed that two samples of C and S-type MBAs present two different values for the transition diameter to a Maxwellian distribution of the rotation frequency, respectively $44 \pm 2$ and $30 \pm 1$ km (Carbognani, 2011). On account of this difference, and using a YORP (Yarkovsky-O'Keefe-Radzievskii-Paddack) effect model, we made an independent estimate of the ratio between the bulk densities of S and C-type asteroids giving $\rho_S/\rho_C \approx 2.9 \pm 0.3$ (Carbognani, 2011). The value, as obtained, is representative of asteroids with small diameter, between 30 to 50 km, and so it seems compatible with the expected low bulk density for the small C-type asteroids. With this ratio we have:

$$\left(\frac{P_C}{P_S}\right)_Y \approx 1.7 \pm 0.1 \qquad (3)$$

In the paper we also keep in mind the value of Eq. (3), that may be useful for comparison with Eq. (2).

## The selection of the Main Belt sample

To verify the previous predictions, we compare the spin-barrier values of numbered MBAs belonging to the Tholen/SMASSII C and S classes (Tholen, 1989; Bus and Binzel, 2002), with dimensions between 1 and 500 km, not belonging to families or binary systems and with a $U$ quality of the lightcurve equal or higher than 2. A $U$ value equal to 2, means that the lightcurve was not completely observed and that the period uncertainty can arrive up to 30%. The same $U$ value is also used in the cases in which the true period can be a whole multiple of the observed lightcurve.

The data for the MBAs sample are drawn from the asteroid LightCurve Data Base (LCDB), version of 2017 April 3 (Warner, 2009), which contains data on 18,478 numbered asteroids. For asteroids extraction a small software that read the LCDB file and saved the results in a new file was written. In this work only asteroids whose taxonomic classification was made on the basis of color indices or spectrum were considered. In this way we rule out the asteroids for which the taxonomic class is simply

"assumed" by the orbit. The asteroids discarded for this reason are about 89% of the whole C+S-type sample in LCDB (about 7,200 asteroids): unfortunately there are no reliable taxonomic data on many MBAs.

The list obtained in this way includes 774 MBAs (range 1-500 km), of which 388 are S-type and 385 C-type. The plot frequency vs. diameter for this sample is shown in Fig. 2: the presence of the spin-barrier at least in the 4-20 km range is evident. Below 4 km the MBAs with a known rotation period and a sure taxonomic class are very few, while over 20 km there is a net departure from the spin-barrier constant value.

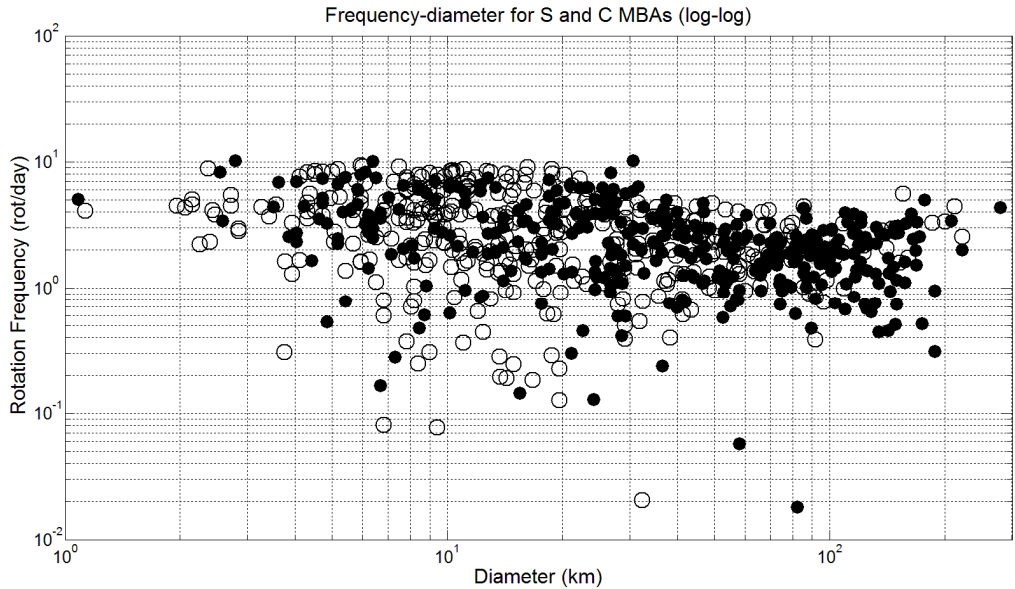

Figure 2. The frequency-diameter plot for our MBAs sample whose taxonomic classification was made on the basis of color indices or spectrum were considered (C-type black dot, S-type white dot).

An enlargement of the region between 4 and 20 km of Fig. 2 shows that the maximum rotation frequency is about 10 rot/day and that the number of S-type asteroids is higher than C (see Fig. 3). This last difference is due to the greater ease of observability for S-type asteroids, so it is an observational bias. However all diameters appear well represented, there are no obvious gaps.

It is interesting to note that if one looks at the plot of all known spin rates, including MBAs object for which the taxonomic class (C and S-type), is simply assumed by the orbit, the trend of the rotational frequency between 4 and 20 km changes. The maximum value of the rotational frequency rises from 10 to 12 (see Fig. 4 and compare with Fig. 3). This means that there are several high-frequency rotation MBAs that do not have a secure taxonomic classification and that we can not include in our analysis. However in this paper we are particularly interested in the ratio between the spin-barrier of C and S-type asteroids, and not in their absolute values. With the discussion about Eq. (4) and (5) we shown that it is reasonable to expect that our results are not too influenced by the bias of the sample. When the data on the taxonomic classification of MBAs will be more numerous we can also make a reasonable estimate of the asteroids bulk density starting from the absolute value of the spin-barrier. Now we want to move on to analyze the ratio between the spin-barrier values for S and C-type considering the asteroids with high rotation frequencies only.

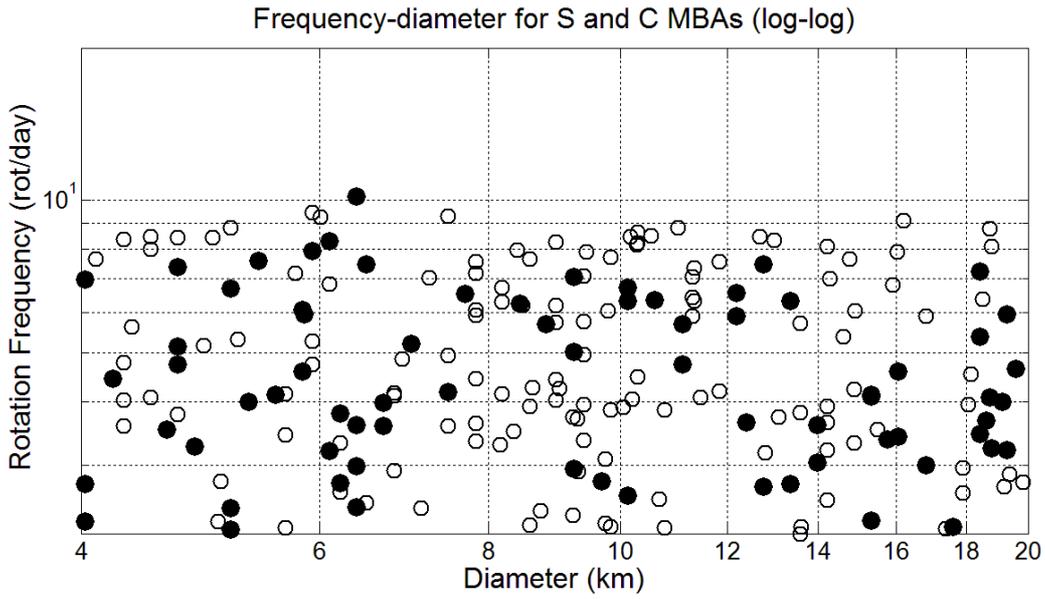

Figure 3. Magnification of the diameter range between 4 and 20 km for our sample of Fig. 2. The maximum rotation frequency is about 10 rot/day (C-type black dot, S-type white dot). From this figure we can see that S-type asteroids tend to be more higher than C-type.

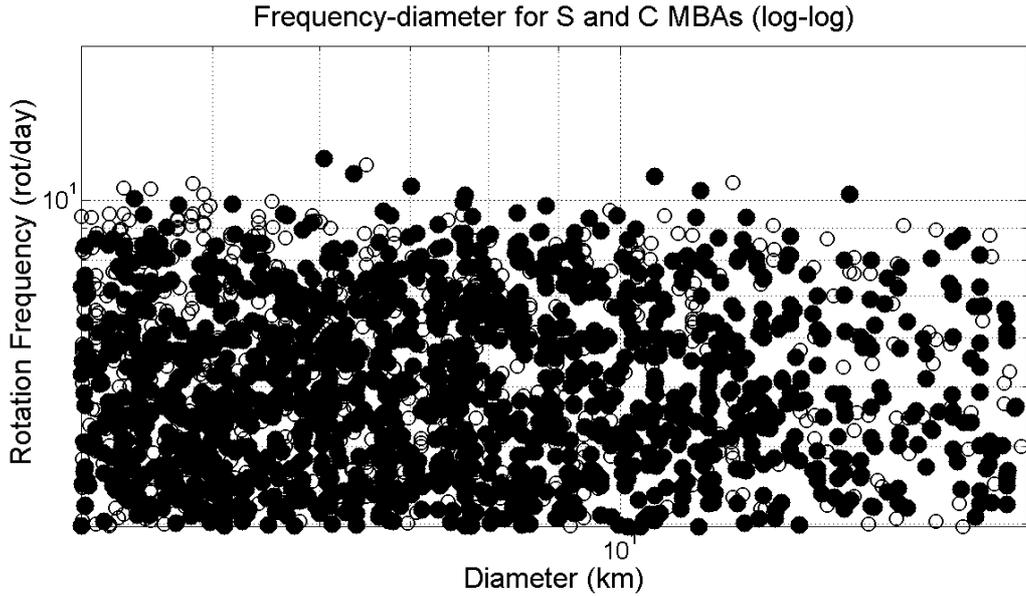

Figure 4. The same region, with diameter between 4 and 20 km, of Fig. 3 but also including MBAs object (C and S-type), for which the taxonomic class is simply assumed by the orbit. The number of asteroids is greater and the maximum rotation frequency rise to about 12 rot/day.

## Estimate of the spin-barriers ratio for C and S-type asteroids

The first step of our analysis consists in determining the spin-barrier value for C and S asteroids. For this purpose, in the 4-20 km range, we manually selected the asteroid with the highest rotation frequency in such a way to take the "external envelope" of the plot of Fig.3. In this way we selected 32 S-type and 20 C-type asteroids (see Fig.5). For C-type the mean spin-barrier value proves to be $6.9 \pm 0.2$ rot/day ($P_C \approx 3.5$ h), while for S-type is $8.2 \pm 0.1$ rot/day ($P_S \approx 3$ h). The uncertainty is the mean standard deviation. With these spin-barrier frequencies, from Eq. (1), the corresponding bulk densities are $0.90 \pm 0.05$ g/cm$^3$ for C-type and 1.27 ± 0.04 g/cm³ for S-type. These values are somewhat lower than the best bulk density estimates which we reported in the introduction and they seem unrealistic, especially for S-type asteroids (for C-type asteroids the value of 0.9 g/cm³ is within the uncertainty). However, in this work we are not interested with the spin-barrier absolute values but with their ratio and this is not much sensitive to the fact that we take the mean or the highest values of the rotational frequencies. To prove this consider the ratio between the mean rotation frequency plus a quantity $\Delta\omega \approx 1$ that shift from the mean value to the highest observed values. In this case we can write the spin-barrier ratio as follow:

$$\frac{\omega_S}{\omega_C} = \frac{\omega_{0S} + \Delta\omega_S}{\omega_{0C} + \Delta\omega_C} = \frac{\omega_{0S}}{\omega_{0C}}\left(1 + \frac{\Delta\omega_S}{\omega_{0S}}\right)\Big/\left(1 + \frac{\Delta\omega_C}{\omega_{0C}}\right) \qquad (4)$$

In our case we have $\Delta\omega_S/\omega_{0S} \approx 0.12$ and $\Delta\omega_C/\omega_{0C} \approx 0.14$ and:

$$\frac{\omega_S}{\omega_C} \approx 0.98\,\frac{\omega_{0S}}{\omega_{0C}} \qquad (5)$$

So the ratio would not change considerably even if we take into accout the highest rotation frequencies, but the mean value is more representative of the behavior of the considered diameters range. Bearing in mind this fact the ratio between the C and S spin-barrier (mean values) result about 1.20 ± 0.04, compatible from the expected values given by Eq. (2) but not by Eq. (3). Also Chan-Kao et al. (2015), using a different method and asteroids sample, found a difference in the spin-rate distribution for C and S-type MBAs. In this work, numeric values are not explicitly provided but from Fig. 12 of Chan-Chao et al., we have $\omega_S \approx 10$ rot/day and $\omega_C \approx 8$ rot/day, i.e. $\omega_S/\omega_C \approx 1.25$ in good agreement with our findings.

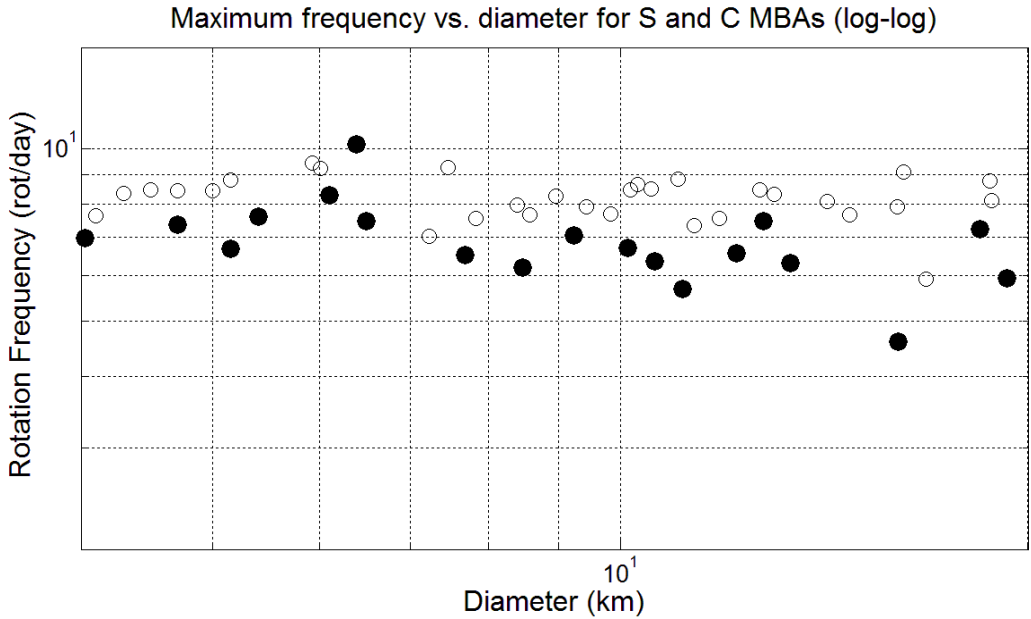

Figure 5. The high frequency-diameter plot in the range 4-20 km. The asteroids have been manually selected in such a way to take the "external envelope" of the plot in Fig.3. In an ideal world asteroids should all be aligned to trace the same curve. In the real world every asteroid is a case in itself and there are observational bias (C-type black circle, S-type white circle).

If we limit ourselves to the 4-10 km diameter interval, there are 10 C-type asteroids and 16 S-type asteroids. In this case for C-type the mean spin-barrier value rise to 7.4 ± 0.3 rot/day ($P_C \approx 3.2$ h), while for S-type is still 8.2 ± 0.2 rot/day. So the ratio between the C and S spin-barrier values drop to about 1.11 ± 0.05. This result was predictable by looking at the distribution of C and S-type objects, both in Fig. 3 and Fig. 5, in which the asteroids of the two classes tend to overlap while the diameter decrease. The limit value of about 10 km is interesting because it is below this value that the effects of the cohesive strength, between the blocks of which a rubble-pile asteroid is composed, would be sensitive (Sánchez and Scheeres, 2014).

To further verify whether the observed spin-barriers of C and S-type asteroids have something in common or not we have chosen to compare the cumulative distributions of all asteroids (not just those with the highest rotation frequency),  in the range 4-20 km (see Fig. 6). In this way there are, at least, 100 asteroids for each taxonomic class. This limit is important to have a reasonable number of asteroids on which to apply statistical analysis. The cumulative distribution presents the asteroids number with value of the independent variable (in this case the rotation frequency), under a certain threshold. To check the compatibility we have used the Kolmogorov-Smirnov (KS) test (Press et al., 1992). The result of the test shows that the two rotation frequency distribution are different with a probability of about 70%. This value rise up to 95% if we consider also asteroids that do not have a secure taxonomic classification.

In the range 4-10 km instead the KS test tell us that the frequency distributions are equal with a probability of about 85%. Even with all the limits of the statistical tests, the results of the KS are consistent with what was found by computing the simple mean values of the maximum rotational frequencies: that there is a difference between the spin-barrier ratio of C-type and S-type asteroids in the range 4-20 km and 4-10 km.

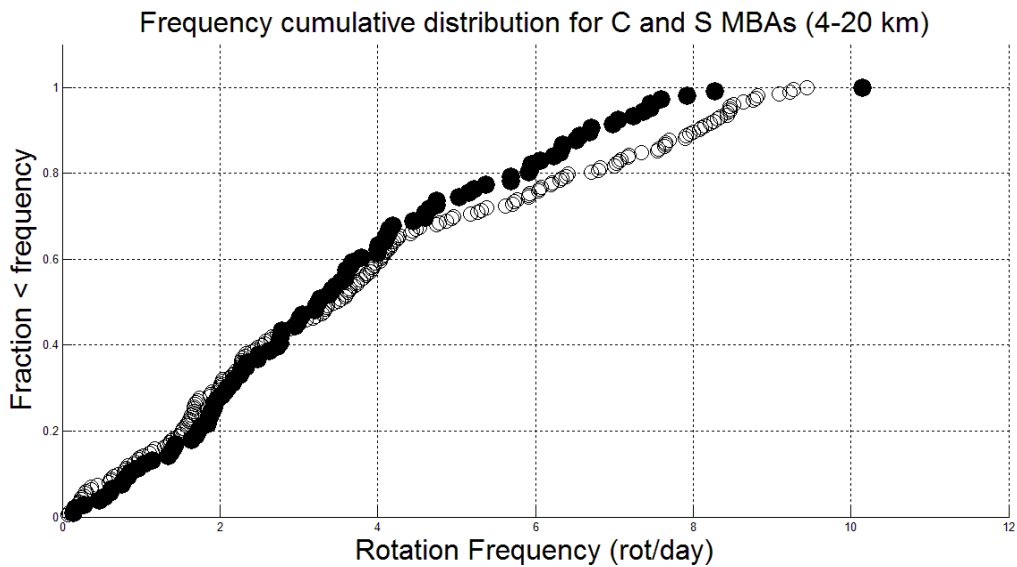

Figure 6. The frequency cumulative distributions of our MBAs sample in the range 4-20 km. There are 106 C-type and 203 S-type asteroids (C-type black circle, S-type white circle).

## The role of the shape as regard the spin-barrier ratio

In this section we want to estimate the importance of the shape as regard the spin-barrier ratio given by Eq. (2). The asteroids are not spherical bodies as supposed in Eq. (1), instead they are much more similar to triaxial ellipsoids. For a triaxial ellipsoid the spin-barrier formula is more complex: it appears a new numerical coefficient, that multiplies Eq. (1), that takes into account how the ellipsoid is elongated (Pravec and Harris, 2000; Richardson et al., 2005 for more details):

$$P_{crit} \approx \frac{3.3 \, h}{\sqrt{\rho}} \sqrt{1 + \Delta m} \qquad (6)$$

In Eq. (6) $\Delta m$ is the maximum amplitude of light curve variation. We will call this new factor "ellipticity coefficient". This analysis was performed on the same asteroids of Fig. 5, i.e. bodies in the diameter range 4-20 km and with the highest rotation frequency. In order to make the lightcurve amplitudes comparable each other our sample was randomized with the Binzel and Sauter (1992) method. The resulting mean amplitude for C-type is 0.29 ± 0.05 mag, while for S-type is 0.21 ± 0.02 mag. These mean amplitudes give a ratio between the ellipticity coefficients of C-type and S-type equal to about 1.03. We conclude that the randomized amplitudes, from which one can estimate the ellipticity coefficient of Eq. (6), are not so different to change the bulk density ratio given by Eq. (2).

## An estimate of the cohesive strength ratio of C and S-type asteroids

As we have seen, if we limit the spin-barrier comparison between the 4-10 km range diameter the spin-barrier ratio appears lower with a value of about 1.11 ± 0.05. Unfortunately in this range of diameters the representative asteroids are really few, only 16 for S-type and 10 for C-type, and what follows must be taken with caution.

This minor difference between the spin-barrier of C and S-type might be due to the presence of small Van der Waals forces between the different blocks that form the asteroid, as theorized by Holsapple (2007) and further developed Sánchez and Scheeres (2014). These forces give rise to a cohesive strength invoked to explain the existence of the Large Super-Fast Rotators, as 2001 OE84 (Pravec et al., 2002), i.e. asteroids with a diameter larger than 0.15 km and rotation period much lower than 2.2 h. The effects of the cohesive strength would be sensitive up to about 10 km in diameter (Sánchez and Scheeres, 2014).

Now we will check if the previous hypothesis is physically sensible. The spin-barrier rate, in case of asteroid with cohesive strength, is given by (Sánchez and Scheeres, 2014):

$$\omega^2 = \omega_0^2 + \frac{\sigma_Y}{\rho a^2} \qquad (7)$$

In Eq. (7) $\omega_0$ is the cohesionless spin-barrier rotation frequency, $\sigma_Y$ is the cohesive strength (force for unit area i.e. a pressure), and $a$ is the asteroid radius. Of course if $\sigma_Y = 0$ Eq. (7) reduces to Eq. (1). If the asteroids are relatively large (i.e. ≥ 4 km as in our case) and the cohesion is low, the corrective term is small compared to $\omega_0$ and we can write:

$$\omega \approx \omega_0 \left( 1 + \frac{1}{2} \frac{\sigma_Y}{\rho \omega_0^2 a^2} \right) \qquad (8)$$

So, in this approximation, the cohesive strength is an additive term of Eq. (1). From Eq. (8), if we impose that the spin-barrier values are about the same for C and S-type asteroids, and assume that $\omega_{0S}/\omega_{0C} \approx 1.4$ and $\rho_S/\rho_C \approx 2$, we have:

$$0.6 \frac{\omega_{0S}^2 a_S^2 \rho_S}{\sigma_{YS}} \approx \left( 2 \frac{\sigma_{YC}}{\sigma_{YS}} - 1 \right) \qquad (9)$$

Now we assume as physically acceptable values for the quantity on the left side of Eq. (9): $a_S \approx 2.5 \cdot 10^3$ m, $\omega_{0S} \approx 8.2$ rot/day = $9.5 \cdot 10^{-5}$ rot/s, $\rho_S \approx 2700$ kg/m$^3$ and $\sigma_{YS} \approx 50$ Pa (weak cohesion). With these values, the first member of Eq. (9) becomes about 1.8 and we get the following ratio:

$$\frac{\sigma_{YC}}{\sigma_{YS}} \approx 1.4 \qquad (10)$$

In conclusion, if the C-type cohesive strength is higher than about 40% when compared to the S-type value, the spin-barrier of the two populations can be very similar. This request seems reasonable, but of course this is a preliminary result that could change radically if the number of asteroids with known taxonomic class increases in the future.

## Conclusions

We considered a sample of 774 S and C-type MBAs taken by LCDB with known rotation periods and known taxonomic class. Considering the ratio between the bulk densities, the expected ratio between the spin-barrier of C and S-type asteroids, in the cohesionless rubble-pile model, is about $P_C/P_S \approx 1.4 \pm 0.3$. Instead, using a bulk density ratio derived from a YORP effect model the ratio is $P_C/P_S \approx 1.7 \pm 0.1$. In order to verify these predictions we analyzed a MBAs sub-sample with the highest rotation frequency in the 4-20 km range looking for a difference in the spin barrier value and we found $P_C/P_S \approx 1.20 \pm 0.04$. This result cannot be attributed to a difference in the elongated shape of the two different asteroids populations so it appears to be consistent with what was expected. However, in the diameter 4-10 km range the spin barrier ratio seems nearer to 1. A reasonable hypothesis is that it is an effect of a different weak cohesive strength for C and S-type asteroids. To further develop the analysis, and to get the ratio of the spin-barrier vs. the diameter, it is imperative to obtain the taxonomic classification of a larger number of MBAs, especially for dimensions below 20 km where the presence of the spin barrier is evident.

## Acknowledgements

The Astronomical Observatory of the Autonomous Region of the Aosta Valley (OAVdA) is managed by the Fondazione Clément Fillietroz-ONLUS, which is supported by the Regional Government of the Aosta Valley, the Town Municipality of Nus and the "Unité des Communes valdôtaines Mont-Émilius". The research was partially funded by a 2016 "Research and Education" grant from Fondazione CRT. Research at OAVdA was also supported by the 2013 Shoemaker NEO Grant and made use of the NASA's Astrophysics Data System. Many thanks to Alan Harris that with his advice have greatly contributed to the improvement of the work.